\documentstyle[12pt]{article}
\newcounter{eq}
\newcounter{sc}

\def\overleftrightarrow#1{\vbox{\ialign{##\crcr
 $\leftrightarrow$\crcr\noalign{\kern-1pt\nointerlineskip}
 $\hfil\displaystyle{#1}\hfil$\crcr}}}

\newlength{\minitwocolumn}
\begin{document}

%%%%%%%%%%%%%%%%%%%%%%%%%%%%%%%%%%%%%%%%%%%%%%%%%%%%%%%%%%%%%%%%%%
%%%%%%%%%%%%%%%%%%%%%%%% Title %%%%%%%%%%%%%%%%%%%%%%%%%%%%%%%%%%%
%%%%%%%%%%%%%%%%%%%%%%%%%%%%%%%%%%%%%%%%%%%%%%%%%%%%%%%%%%%%%%%%%%
\begin{flushright}
DPUR/TH/54\\
April, 2017\\
\end{flushright}
\vspace{20pt}

%\magnification=\magstep1
\pagestyle{empty}
\baselineskip15pt
%\font\cmssB=cmss17
%\font\cmssS=cmss10

\begin{center}
{\large\bf Manifestly Local Formulation of Nonlocal Approach to the Cosmological Constant Problem
\vskip 1mm }

\vspace{10mm}
Ichiro Oda \footnote{E-mail address:\ ioda@phys.u-ryukyu.ac.jp
}

\vspace{3mm}
           Department of Physics, Faculty of Science, University of the 
           Ryukyus,\\
           Nishihara, Okinawa 903-0213, Japan.\\

\end{center}

%\maketitle

\vspace{3mm}
\begin{abstract}
We present a manifestly local and general coordinate invariant formulation of a nonlocal approach
to the cosmological constant problem which has been recently proposed by Carroll and Remmen.
To do that, we need to introduce a topological term involving a new 3-form gauge field. The equations
of motion for this new 3-form gauge field lead to a constant Lagrange multiplier parameter and
the resulting action becomes equivalent to that of Carroll and Remmen. In our formulation,
nonlocal informations are encoded via the procedure of taking the space-time average at the stage
of the equations of motion. Consequently, our theory evades a no-go theorem by Weinberg 
and provides a new solution to the cosmological constant problem in almost exactly the same way as 
the original proposal by Carroll et al.
\end{abstract}

\newpage
\pagestyle{plain}
\pagenumbering{arabic}
%\setcounter{page}{1}

%%%%%%%%%%%%%%%%%%%%%%%%%%%%%%%%%%%%%%%%%%%%%%%%%%%%%%%%%%%%%%%%%%
%%%%%%%%%%%%%%%%%%%%%%%% Article %%%%%%%%%%%%%%%%%%%%%%%%%%%%%%%%%
%%%%%%%%%%%%%%%%%%%%%%%%%%%%%%%%%%%%%%%%%%%%%%%%%%%%%%%%%%%%%%%%%%

\rm
%%%%%%%%%%%%%%%%%%%%%%%%%%%%%%%%%%%%%%%%%%%%%%%%%%%%%%%%%%%%%%%%%%%%%
%%%%%%%%%%%%%%%%%%%%%%%%%%%%%%   SEC  1    %%%%%%%%%%%%%%%%%%%%%%%%%%
%%%%%%%%%%%%%%%%%%%%%%%%%%%%%%%%%%%%%%%%%%%%%%%%%%%%%%%%%%%%%%%%%%%%%
\section{Introduction}

The cosmological constant problem, that is, why the vacuum energy density or cosmological constant is extremely small
from particle physics standards,  is now considered as one of the central challenges in the elementary particle physics 
and quantum gravity \cite{Weinberg}. It is likely that the ultimate solution to this problem might require a paradigm change of 
currently recoginized fundamental principles in modern theoretical physics.

It is worthwhile to recall the reason why it is difficult to solve the cosmological constant problem within our 
present state of knowledge. First, the problem is related to not classical gravity but quantum gravity since one can 
classically set the cosmological constant to zero by hand without any conflict with fundamental principles of general relativity 
as Einstein indeed did. Quantum mechanically, however, we are not allowed to simply put the cosmological constant
to zero in a microscopic Lagrangian because the cosmological constant receives a lot of contributions from radiative
corrections.  Second, solving this problem seems to need a low energy mechanism to cancel or reduce huge contributions
from radiative corrections, but there is at least currently no such a low energy mechanism. Moreover, it is very hard to
modify the low energy framework in a sensible manner, given all of the familiar successes of particle physics and 
cosmology. 

Faced with this difficulty, one way out is to alter some fundamental principles. In particular, locality might be a natural 
target to be attacked since a no-go theorem by Weinberg \cite{Weinberg} implies that no local field equations can have
a flat Minkowski metric for generic values of parameters, or to put differently, any solution to the cosmological constant 
problem cannot be obtained without fine-tuning within the framework of local field theories. However, to sacrifice the 
principle of locality at the Lagrangian level could lead to violation of unitarity etc. It is therefore plausible to keep the
locality principle at the action level and introduce some nonlocal effects in the equations of motion.

Recently, there has appeared an interesting nonlocal approach to the cosmological constant problem \cite{Carroll}.
In this nonlocal approach, the constraint such that the total action is vanishing plays an important role, and
this constraint is gained via variational principle with respect to a constant Lagrange multiplier parameter.  Although in this
nonlocal approach there does not seem to be violation of causality and unitarity etc. and consistent with all 
cosmological and experimental observations, the presence of the constant  Lagrange multiplier parameter
appears to conflict with some expectation about the macroscopic origin of the deeper mechanism behind this
approach.  In other words, while this nonlocal approach is somewhat ad hoc and phenomenological, it may be serve as a pointer
toward a more comprehensive theory \cite{Carroll}.

In the present article, we attempt to construct a manifestly local and generally coordinate invariant formalism for this nonlocal approach 
to the cosmological constant problem. In our theory, the constant Lagrange multiplier parameter is replaced with a local scalar field 
and the constant Lagrange multiplier parameter is obtained as a classical solution to the equations of motion for a newly
introduced 3-form field. The resulting theory is very similar to that by Carroll et al., but there are some differences between 
the two theories.  

This paper is organised as follows: In Section 2, we propose a manifestly local and generally coordinate invariant action
which reduces to the action by Carroll et al. after taking the variation with respect to a new 3-form gauge
field. In Section 3, all the equations of motion are derived from our action step by step. Taking the space-time average of
the gravitational field equations yields very similar equations to those obtained by Carroll et al.
The final section is devoted to discussions.

%%%%%%%%%%%%%%%%%%%%%%%%%%%%%%%%%%%%%%%%%%%%%%%%%%%%%%%%%%%%%%%%%%%%%
%%%%%%%%%%%%%%%%%%%%%%%%%%%%%%   SEC  2    %%%%%%%%%%%%%%%%%%%%%%%%%%
%%%%%%%%%%%%%%%%%%%%%%%%%%%%%%%%%%%%%%%%%%%%%%%%%%%%%%%%%%%%%%%%%%%%%
\section{Manifestly local formulation}

In this section we wish to present a manifestly local and generally coordinate invariant action for a nonlocal approach to the cosmological 
constant problem by Carroll and Remmen \cite{Carroll}.  Since we would like to preserve the main feature of the original work
by Carroll et al.  \cite{Carroll}, the sector which we add should not gravitate directly. The unique and nontrivial possibility would be to add 
a topological sector to the Carroll and Remmen action. The idea to follow has been already suggested in the gauge invariant formulation of 
unimodular gravity \cite{Einstein, Weinberg} by Henneaux and Teitelboim \cite{Henneaux}. \footnote{We have also used this idea to construct 
topological induced gravity \cite{Oda1}-\cite{Oda3}.}  In the Ref. \cite{Kaloper2}, Kaloper et al. have also generalized the idea by Henneaux 
and Teitelboim to construct a manifestly local, diffeomorphism invariant, and locally Poincar\'e invariant formulation of vacuum energy 
sequestering \cite{Kaloper1, Kaloper3}. In this article, we will in essence follow this generalized idea.

Let us start by presenting a manifestly local and generally coordinate invariant action for the Carroll-Remmen action \footnote{We follow notation 
and conventions of the textbook by Misner et al \cite{MTW}.} 
%**   Our Action   %%%%%%%%%%%%%%%%%%%%%%%%%%%%%%%%%%%%%%%%%%%%%%%%%%%%%%%%%
\begin{eqnarray}
S = S_{CR} + S_{Top},
\label{Our Action}
\end{eqnarray}
%%%%%%%%%%%%%%%%%%%%%%%%%%%%%%%%%%%%%%%%%%%%%%%%%%%%%%%%%%%%%%%%%%%
where the total action consists of two parts, the Carroll-Remmen action $S_{CR}$ \footnote{The last boundary term is included
to obtain stationary action under variations that leave the field strength $F_{\mu\nu\rho\sigma}$ fixed on the boundary \cite{Duncan}.} 
and the topological action $S_{Top}$,  
and each action in the RHS is concretely defined as follows:
%**   CR Action   %%%%%%%%%%%%%%%%%%%%%%%%%%%%%%%%%%%%%%%%%%%%%%%%%%%%%%%%%
\begin{eqnarray}
S_{CR}  = \int d^4 x \sqrt{- g} \ \eta(x) \left[ \frac{1}{16 \pi G}  ( R - 2 \Lambda ) + {\cal{L}}_m
- \frac{1}{2} \cdot \frac{1}{4!} F_{\mu\nu\rho\sigma}^2 + \frac{1}{6} \nabla_\mu ( F^{\mu\nu\rho\sigma} A_{\nu\rho\sigma} )
\right],
\label{CR Action}
\end{eqnarray}
%%%%%%%%%%%%%%%%%%%%%%%%%%%%%%%%%%%%%%%%%%%%%%%%%%%%%%%%%%%%%%%%%%%
and
%**   Top Action   %%%%%%%%%%%%%%%%%%%%%%%%%%%%%%%%%%%%%%%%%%%%%%%%%%%%%%%%%
\begin{eqnarray}
S_{Top}  = \int d^4 x  \ \frac{1}{4!}  \ \mathring{\varepsilon}^{\mu\nu\rho\sigma} \eta(x) H_{\mu\nu\rho\sigma}.
\label{Top Action}
\end{eqnarray}
%%%%%%%%%%%%%%%%%%%%%%%%%%%%%%%%%%%%%%%%%%%%%%%%%%%%%%%%%%%%%%%%%%%
Here we will account for various quantities: $g$ is the determinant of the metric tensor, $g = \det g_{\mu\nu}$, and $R$ denotes the
scalar curvature. $\eta(x)$ is a scalar field, which reduces to the constant Lagrange multiplier parameter of Carroll et al. after taking a classical
solution. $G$, $\Lambda$, and $ {\cal{L}}_m$ are the Newton constant, the bare cosmological constant and the Lagrangian density for
generic matter fields, respectively. Moreover, $F_{\mu\nu\rho\sigma}$ and $ H_{\mu\nu\rho\sigma}$ are respectively the field strengths 
for two 3-form gauge fields $A_{\mu\nu\rho}$ and $B_{\mu\nu\rho}$, those are, 
%**   4-forms   %%%%%%%%%%%%%%%%%%%%%%%%%%%%%%%%%%%%%%%%%%%%%%%%%%%%%%%%%
\begin{eqnarray}
F_{\mu\nu\rho\sigma} = 4 \partial_{[\mu} A_{\nu\rho\sigma]},  \qquad 
H_{\mu\nu\rho\sigma} = 4 \partial_{[\mu} B_{\nu\rho\sigma]},
\label{4-forms}
\end{eqnarray}
%%%%%%%%%%%%%%%%%%%%%%%%%%%%%%%%%%%%%%%%%%%%%%%%%%%%%%%%%%%%%%%%%%%
where the square brackets denote antisymmetrization of enclosed indices. Finally, $\mathring{\varepsilon}^{\mu\nu\rho\sigma}$
and $\mathring{\varepsilon}_{\mu\nu\rho\sigma}$ are the Levi-Civita tensor density defined as
%**   Levi   %%%%%%%%%%%%%%%%%%%%%%%%%%%%%%%%%%%%%%%%%%%%%%%%%%%%%%%%%
\begin{eqnarray}
\mathring{\varepsilon}^{0123} = + 1,  \qquad 
\mathring{\varepsilon}_{0123} = - 1,
\label{Levi}
\end{eqnarray}
%%%%%%%%%%%%%%%%%%%%%%%%%%%%%%%%%%%%%%%%%%%%%%%%%%%%%%%%%%%%%%%%%%%
and they are related to the totally antisymmetric tensors $\varepsilon^{\mu\nu\rho\sigma}$ and $\varepsilon_{\mu\nu\rho\sigma}$ via
relations
%**   Levi-tensor  %%%%%%%%%%%%%%%%%%%%%%%%%%%%%%%%%%%%%%%%%%%%%%%%%%%%%%%%%
\begin{eqnarray}
\varepsilon^{\mu\nu\rho\sigma} = \frac{1}{\sqrt{-g}}  \mathring{\varepsilon}^{\mu\nu\rho\sigma},  \qquad 
\varepsilon_{\mu\nu\rho\sigma} = \sqrt{-g}  \mathring{\varepsilon}_{\mu\nu\rho\sigma}.
\label{Levi-tensor}
\end{eqnarray}
%%%%%%%%%%%%%%%%%%%%%%%%%%%%%%%%%%%%%%%%%%%%%%%%%%%%%%%%%%%%%%%%%%%
Also note that the Levi-Civita tensor density satisfies the following equations:
%**   e-identity  %%%%%%%%%%%%%%%%%%%%%%%%%%%%%%%%%%%%%%%%%%%%%%%%%%%%%%%%%
\begin{eqnarray}
\mathring{\varepsilon}^{\mu\nu\rho\sigma} \mathring{\varepsilon}_{\alpha\beta\rho\sigma} = - 2 ( \delta_\alpha^\mu \delta_\beta^\nu
-   \delta_\alpha^\nu \delta_\beta^\mu ),  \qquad 
\mathring{\varepsilon}^{\mu\nu\rho\sigma} \mathring{\varepsilon}_{\alpha\nu\rho\sigma} = - 3!  \delta_\alpha^\mu,  \qquad 
\mathring{\varepsilon}^{\mu\nu\rho\sigma} \mathring{\varepsilon}_{\mu\nu\rho\sigma} = - 4!.
\label{e-identity}
\end{eqnarray}
%%%%%%%%%%%%%%%%%%%%%%%%%%%%%%%%%%%%%%%%%%%%%%%%%%%%%%%%%%%%%%%%%%%

At this stage, it is worth commenting on some features of the topological action $S_{Top}$ in (\ref{Top Action}). This term is
free from the metric tensor so that its metric variation is identically vanishing, thereby implying that it does not gravitate directly.
However, the existence of the scalar field $\eta(x)$ in the both  (\ref{CR Action}) and  (\ref{Top Action}) makes it possible
to link the 4-form field strength $H_{\mu\nu\rho\sigma}$ in $S_{Top}$ to the geometry, i.e., the scalar curvature $R$, and matter fields in
$S_{CR}$. In other words, the topological sector {\it sees} information on cosmological constant indirectly as will be found later. 

We will now show that our action becomes equivalent to that of Carroll et al. To do that, let us take the variation with respect to
the 3-form $B_{\mu\nu\rho}$, which gives us the equations for a scalar field $\eta(x)$
%**   eta-eq  %%%%%%%%%%%%%%%%%%%%%%%%%%%%%%%%%%%%%%%%%%%%%%%%%%%%%%%%%
\begin{eqnarray}
\mathring{\varepsilon}^{\mu\nu\rho\sigma} \partial_\sigma \eta(x)  = 0,
\label{eta-eq}
\end{eqnarray}
%%%%%%%%%%%%%%%%%%%%%%%%%%%%%%%%%%%%%%%%%%%%%%%%%%%%%%%%%%%%%%%%%%%
from which we have a classical solution for $\eta(x)$
%**   eta-sol  %%%%%%%%%%%%%%%%%%%%%%%%%%%%%%%%%%%%%%%%%%%%%%%%%%%%%%%%%
\begin{eqnarray}
\eta(x)  = \eta,
\label{eta-sol}
\end{eqnarray}
%%%%%%%%%%%%%%%%%%%%%%%%%%%%%%%%%%%%%%%%%%%%%%%%%%%%%%%%%%%%%%%%%%%
where $\eta$ is a certain constant. Substituting this solution into the starting action (\ref{Our Action}),
we arrive at the action proposed by Carroll et al. to solve the cosmological constant problem \cite{Carroll}:
%**   Original CR Action   %%%%%%%%%%%%%%%%%%%%%%%%%%%%%%%%%%%%%%%%%%%%%%%%%%%%%%%%%
\begin{eqnarray}
S  = \eta \int d^4 x \sqrt{- g} \left[ \frac{1}{16 \pi G}  ( R - 2 \Lambda ) + {\cal{L}}_m
- \frac{1}{2} \cdot \frac{1}{4!} F_{\mu\nu\rho\sigma}^2 + \frac{1}{6} \nabla_\mu ( F^{\mu\nu\rho\sigma} A_{\nu\rho\sigma} )
\right].
\label{Original CR Action}
\end{eqnarray}
%%%%%%%%%%%%%%%%%%%%%%%%%%%%%%%%%%%%%%%%%%%%%%%%%%%%%%%%%%%%%%%%%%%
Note that with Eq. (\ref{eta-sol}), the topological term $S_{Top}$ becomes a surface term, which is ignored in this derivation.

%%%%%%%%%%%%%%%%%%%%%%%%%%%%%%%%%%%%%%%%%%%%%%%%%%%%%%%%%%%%%%%%%%%%%
%%%%%%%%%%%%%%%%%%%%%%%%%%%%%%   SEC  3    %%%%%%%%%%%%%%%%%%%%%%%%%%
%%%%%%%%%%%%%%%%%%%%%%%%%%%%%%%%%%%%%%%%%%%%%%%%%%%%%%%%%%%%%%%%%%%%%
\section{Equations of motion}

Let us derive the equations of motion from the action (\ref{Our Action}) in order which are completely local. 
Now the variation of the action (\ref{Our Action}) with respect to the scalar field $\eta(x)$ provides us with the vanishing
total Lagrangian density
%**   Zero-Lagr   %%%%%%%%%%%%%%%%%%%%%%%%%%%%%%%%%%%%%%%%%%%%%%%%%%%%%%%%%
\begin{eqnarray}
\sqrt{- g} \left[ \frac{1}{16 \pi G}  ( R - 2 \Lambda ) + {\cal{L}}_m
- \frac{1}{2} \cdot \frac{1}{4!} F_{\mu\nu\rho\sigma}^2 + \frac{1}{6} \nabla_\mu ( F^{\mu\nu\rho\sigma} A_{\nu\rho\sigma} )
\right] + \frac{1}{4!}  \ \mathring{\varepsilon}^{\mu\nu\rho\sigma} H_{\mu\nu\rho\sigma} = 0.
\label{Zero-Lagr}
\end{eqnarray}
%%%%%%%%%%%%%%%%%%%%%%%%%%%%%%%%%%%%%%%%%%%%%%%%%%%%%%%%%%%%%%%%%%%
Next, we set $H_{\mu\nu\rho\sigma}$ to be
%**   H   %%%%%%%%%%%%%%%%%%%%%%%%%%%%%%%%%%%%%%%%%%%%%%%%%%%%%%%%%
\begin{eqnarray}
H_{\mu\nu\rho\sigma}  = c(x) \varepsilon_{\mu\nu\rho\sigma} = c(x) \sqrt{-g} \mathring{\varepsilon}_{\mu\nu\rho\sigma},
\label{H}
\end{eqnarray}
%%%%%%%%%%%%%%%%%%%%%%%%%%%%%%%%%%%%%%%%%%%%%%%%%%%%%%%%%%%%%%%%%%%
where $c(x)$ is some scalar function. Then, using the relation in Eq. (\ref{e-identity}), Eq.  (\ref{Zero-Lagr}) can be rewritten as
%**   Zero-Lagr 2   %%%%%%%%%%%%%%%%%%%%%%%%%%%%%%%%%%%%%%%%%%%%%%%%%%%%%%%%%
\begin{eqnarray}
\frac{1}{16 \pi G}  \left[  R - 2 \Lambda(x) \right] + {\cal{L}}_m
- \frac{1}{2} \cdot \frac{1}{4!} F_{\mu\nu\rho\sigma}^2 + \frac{1}{6} \nabla_\mu ( F^{\mu\nu\rho\sigma} A_{\nu\rho\sigma} )
= 0,
\label{Zero-Lagr 2}
\end{eqnarray}
%%%%%%%%%%%%%%%%%%%%%%%%%%%%%%%%%%%%%%%%%%%%%%%%%%%%%%%%%%%%%%%%%%%
where we have defined 
%**   Lambda(x)   %%%%%%%%%%%%%%%%%%%%%%%%%%%%%%%%%%%%%%%%%%%%%%%%%%%%%%%%%
\begin{eqnarray}
\Lambda(x) = \Lambda + 8 \pi G c(x).
\label{Lambda(x)}
\end{eqnarray}
%%%%%%%%%%%%%%%%%%%%%%%%%%%%%%%%%%%%%%%%%%%%%%%%%%%%%%%%%%%%%%%%%%%
The equations of motion for the 3-form $A_{\mu\nu\rho}$ take the form
%**   A-eq   %%%%%%%%%%%%%%%%%%%%%%%%%%%%%%%%%%%%%%%%%%%%%%%%%%%%%%%%%
\begin{eqnarray}
\nabla^\mu F_{\mu\nu\rho\sigma} = 0.
\label{A-eq}
\end{eqnarray}
%%%%%%%%%%%%%%%%%%%%%%%%%%%%%%%%%%%%%%%%%%%%%%%%%%%%%%%%%%%%%%%%%%%
As in $H_{\mu\nu\rho\sigma}$, if we set 
%**   F   %%%%%%%%%%%%%%%%%%%%%%%%%%%%%%%%%%%%%%%%%%%%%%%%%%%%%%%%%
\begin{eqnarray}
F_{\mu\nu\rho\sigma}  = \theta(x) \varepsilon_{\mu\nu\rho\sigma},
\label{F}
\end{eqnarray}
%%%%%%%%%%%%%%%%%%%%%%%%%%%%%%%%%%%%%%%%%%%%%%%%%%%%%%%%%%%%%%%%%%%
with $\theta(x)$ being a scalar function, Eq. (\ref{A-eq}) requires $\theta(x)$ to be a constant
%**   Theta   %%%%%%%%%%%%%%%%%%%%%%%%%%%%%%%%%%%%%%%%%%%%%%%%%%%%%%%%%
\begin{eqnarray}
\theta(x) = \theta,
\label{Theta}
\end{eqnarray}
%%%%%%%%%%%%%%%%%%%%%%%%%%%%%%%%%%%%%%%%%%%%%%%%%%%%%%%%%%%%%%%%%%%
where $\theta$ is a constant. Using this fact, Eq.   (\ref{Zero-Lagr 2}) is further simplified to be
%**   R-eq   %%%%%%%%%%%%%%%%%%%%%%%%%%%%%%%%%%%%%%%%%%%%%%%%%%%%%%%%%
\begin{eqnarray}
\frac{1}{16 \pi G}  \left[  R - 2 \Lambda(x) \right] + {\cal{L}}_m - \frac{1}{2} \theta^2 = 0.
\label{R-eq}
\end{eqnarray}
%%%%%%%%%%%%%%%%%%%%%%%%%%%%%%%%%%%%%%%%%%%%%%%%%%%%%%%%%%%%%%%%%%%

Next, with the help of Eq. (\ref{eta-sol}), the metric variation gives the standard gravitational field equations:
%**   Eins-eq 1  %%%%%%%%%%%%%%%%%%%%%%%%%%%%%%%%%%%%%%%%%%%%%%%%%%%%%%%%%
\begin{eqnarray}
\frac{1}{16 \pi G} \left(  G_{\mu\nu}  + \Lambda g_{\mu\nu} \right) - \frac{1}{2} T_{\mu\nu}
+ \frac{1}{4} \cdot \frac{1}{4!} g_{\mu\nu} F_{\alpha\beta\gamma\delta}^2 
- \frac{1}{12} F_{\mu\alpha\beta\gamma} F_\nu \,^{\alpha\beta\gamma}
= 0,
\label{Eins-eq 1}
\end{eqnarray}
%%%%%%%%%%%%%%%%%%%%%%%%%%%%%%%%%%%%%%%%%%%%%%%%%%%%%%%%%%%%%%%%%%%
where $G_{\mu\nu} = R_{\mu\nu} - \frac{1}{2} g_{\mu\nu} R$ is the well-known Einstein tensor and
the energy-momentum tensor is defined by $ T_{\mu\nu} = - \frac{2}{\sqrt{-g}} \frac{\delta (\sqrt{-g}  {\cal{L}}_m)}
{\delta g^{\mu\nu}}$ as usual. Note that we have no contribution from the topological action $S_{Top}$ because
it does not include the metric tensor. Then, using Eqs.  (\ref{F}) and  (\ref{Theta}), Eq.  (\ref{Eins-eq 1}) can be
cast to the form   
%**   Eins-eq   %%%%%%%%%%%%%%%%%%%%%%%%%%%%%%%%%%%%%%%%%%%%%%%%%%%%%%%%%
\begin{eqnarray}
\frac{1}{16 \pi G} \left(  G_{\mu\nu}  + \Lambda g_{\mu\nu} \right) - \frac{1}{2} T_{\mu\nu}
+ \frac{1}{4} \theta^2 g_{\mu\nu} 
= 0,
\label{Eins-eq}
\end{eqnarray}
%%%%%%%%%%%%%%%%%%%%%%%%%%%%%%%%%%%%%%%%%%%%%%%%%%%%%%%%%%%%%%%%%%%
which means that the effective cosmological constant is given by 
%**   Effective CC   %%%%%%%%%%%%%%%%%%%%%%%%%%%%%%%%%%%%%%%%%%%%%%%%%%%%%%%%%
\begin{eqnarray}
\Lambda_{eff} = \Lambda + 4 \pi G \theta^2.
\label{Effective CC}
\end{eqnarray}
%%%%%%%%%%%%%%%%%%%%%%%%%%%%%%%%%%%%%%%%%%%%%%%%%%%%%%%%%%%%%%%%%%%
In this way, we have succeeded in getting two important equations  (\ref{R-eq}) and  (\ref{Eins-eq}).
The two equations must be mathematically consistent in the sense that the trace part of the gravitational field equations
(\ref{Eins-eq}) should be equivalent to Eq. (\ref{R-eq}). It turns out that this consistency condition connects the scalar function $c(x)$
with matter fields, the bare cosmological constant $\Lambda$ and the constant $\theta$:
%**   c(x)   %%%%%%%%%%%%%%%%%%%%%%%%%%%%%%%%%%%%%%%%%%%%%%%%%%%%%%%%%
\begin{eqnarray}
c(x) = {\cal{L}}_m - \frac{1}{2} T + \frac{\Lambda}{8 \pi G} + \frac{1}{2} \theta^2, 
\label{c(x)}
\end{eqnarray}
%%%%%%%%%%%%%%%%%%%%%%%%%%%%%%%%%%%%%%%%%%%%%%%%%%%%%%%%%%%%%%%%%%%
where $T = g^{\mu\nu} T_{\mu\nu}$. This on-shell relation implies that the 4-form field strength $H_{\mu\nu\rho\sigma}$
{\it sees} not only topological information of a background manifold but also information on the cosmological constant existing 
in the Carroll-Remmen action $S_{CR}$. 

Thus far, our theory is completely based on the local field theories, so it is difficult to solve the cosmological constant problem
as shown by Weinberg \cite{Weinberg}. To tackle this hard problem, one way out is to introduce some sorts of nonlocal effects
to our theory. A recipe for getting such nonlocal effects is to take the space-time average of the gravitational field equations in order to
extract information on the bare cosmological constant $\Lambda$. \footnote{The idea of taking the space-time average has
been also used in \cite{Linde, Tseytlin, Nima}.}  Let us therefore trace over Eq.  (\ref{R-eq}) over all
of space-time whose result is given by
%**   Lambda-eq   %%%%%%%%%%%%%%%%%%%%%%%%%%%%%%%%%%%%%%%%%%%%%%%%%%%%%%%%%
\begin{eqnarray}
\Lambda = \frac{1}{2} \langle R \rangle + 8 \pi G \left( \langle \hat {\cal{L}}_m \rangle -  \frac{1}{2} \theta^2 \right), 
\label{Lambda-eq}
\end{eqnarray}
%%%%%%%%%%%%%%%%%%%%%%%%%%%%%%%%%%%%%%%%%%%%%%%%%%%%%%%%%%%%%%%%%%%
where the space-time average of any quantity $A$ is defined as $\langle A \rangle = \frac{\int d^4 x \sqrt{-g} A}{\int d^4 x \sqrt{-g}}$
and we have defined $\hat {\cal{L}}_m = {\cal{L}}_m - c(x)$. To eliminate the bare cosmological constant $\Lambda$ in Eq. (\ref{Eins-eq}),
we will insert Eq.  (\ref{Lambda-eq}) to Eq. (\ref{Eins-eq}). After a straightforward calculation, we obtain
%**   Mod-Eins-eq   %%%%%%%%%%%%%%%%%%%%%%%%%%%%%%%%%%%%%%%%%%%%%%%%%%%%%%%%%
\begin{eqnarray}
G_{\mu\nu}  + \frac{1}{2} \langle R \rangle g_{\mu\nu} = 8 \pi G \left[ T_{\mu\nu} - \langle \hat {\cal{L}}_m \rangle g_{\mu\nu} \right]. 
\label{Mod-Eins-eq}
\end{eqnarray}
%%%%%%%%%%%%%%%%%%%%%%%%%%%%%%%%%%%%%%%%%%%%%%%%%%%%%%%%%%%%%%%%%%%
This equation precisely coincides with the important equation in Ref. \cite{Carroll} except the replacement of ${\cal{L}}_m$ by $\hat {\cal{L}}_m$.
In the present formulation, this replacement is very natural since there is an additional 3-form gauge field in our theory compared to the Carroll 
and Remmen theory. Indeed, it is easy to see that the effective cosmological constant can be read out from Eq. (\ref{Mod-Eins-eq}) to be
%**   Effective CC2   %%%%%%%%%%%%%%%%%%%%%%%%%%%%%%%%%%%%%%%%%%%%%%%%%%%%%%%%%
\begin{eqnarray}
\Lambda_{eff} &=& \langle \frac{1}{2} R + 8 \pi G \hat {\cal{L}}_m \rangle   \nonumber\\
&=& 8 \pi G \langle \frac{1}{2} T - \hat {\cal{L}}_m \rangle  \nonumber\\
&=& \Lambda + 4 \pi G \theta^2,
\label{Effective CC2}
\end{eqnarray}
%%%%%%%%%%%%%%%%%%%%%%%%%%%%%%%%%%%%%%%%%%%%%%%%%%%%%%%%%%%%%%%%%%%
where the last expression is the same in both our theory and the Carroll and Remmen theory, and exactly agrees with Eq.  (\ref{Effective CC}).
As shown in \cite{Carroll}, this effective cosmological constant vanishes for vacuum configuration of the matter fields, and the flat Minkowski
space-time is a classical solution of Eq. (\ref{Mod-Eins-eq}). Thus, our formulation evades the no-go theorem by Weinberg \cite{Weinberg}
by introducing nonlocal effects.

%%%%%%%%%%%%%%%%%%%%%%%%%%%%%%%%%%%%%%%%%%%%%%%%%%%%%%%%%%%%%%%%%%%%%
%%%%%%%%%%%%%%%%%%%%%%%%%%%%%%   SEC  7    %%%%%%%%%%%%%%%%%%%%%%%%%%
%%%%%%%%%%%%%%%%%%%%%%%%%%%%%%%%%%%%%%%%%%%%%%%%%%%%%%%%%%%%%%%%%%%%%
\section{Discussions}

In this article, we have formulated a manifestly local and generally coordinate invariant formulation for the Carroll-Remmen theory,
and clarified the origin of the constant Lagrange multiplier parameter. We have taken into consideration nonlocal effects, i.e., the space-time average, 
at the stage of the equations of motion in order to evade the Weinberg no-go theorem. 

The key idea of our local formulation is to introduce a topological term involving a new 3-form gauge field in addition to the original 3-form
gauge field existing in the Carroll-Remmen action. The latter 3-form gauge field has so far received considerable attention since it contains
no dynamical mode and contributes to the total energy. On the other hand, the former 3-form gauge field has a different role from the latter one
although the former one makes a contribution to the vacuum energy as well.  Namely, the field strength of this new 3-form takes the values dependent 
on space-time coordinates even on-shell, so it knows information on the cosmological constant spread over the space-time. In the Carroll-Remmen theory,
only a constant mode of the new field strength appears in the overall action as a constant Lagrange multiplier parameter.

As an important feature problem, we must understand quantum aspects of our theory. This is a very important step 
for understanding the cosmological constant problem completely.  We wish to consider this problem in near future.

%%%%%%%%%%%%%%%%%%%%%%%%%%%%%%%%%%%%%%%%%%%%%%%%%%%%%%%%%%%%%%%%%%
%%%%%%%%%%%%%%%%%%%%%%%% Acknowledgements %%%%%%%%%%%%%%%%%%%%%%%%%%%%%
%%%%%%%%%%%%%%%%%%%%%%%%%%%%%%%%%%%%%%%%%%%%%%%%%%%%%%%%%%%%%%%%%%
\begin{flushleft}
{\bf Acknowledgements}
\end{flushleft}
This work is supported in part by the Grant-in-Aid for Scientific 
Research (C) No. 16K05327 from the Japan Ministry of Education, Culture, 
Sports, Science and Technology.

%%%%%%%%%%%%%%%%%%%%%%% reference %%%%%%%%%%%%%%%%%%%%%%%%%%%%%%%
%%%%%%%%%%%%%%%%%%%%%%%%%%%%%%%%%%%%%%%%%%%%%%%%%%%%%%%%%%%%%%%%%%

\end{document}